\documentclass[aps,prl,twocolumn,superscriptaddress,showpacs]{revtex4-1}

\usepackage{graphicx}
\usepackage{bm}
\begin{document}

%\preprint{}

\title{Tensor-force effects on single-particle levels
and proton bubble structure around the $Z$ or $N=20$ magic number}


\author{H. Nakada}
\email[E-mail:\,\,]{nakada@faculty.chiba-u.jp}
%\thanks{}
\affiliation{Department of Physics, Graduate School of Science,
 Chiba University\\
Yayoi-cho 1-33, Inage, Chiba 263-8522, Japan}

\author{K. Sugiura}
\affiliation{Department of Physics, Graduate School of Science,
 Chiba University\\
Yayoi-cho 1-33, Inage, Chiba 263-8522, Japan}

\author{J. Margueron}
\affiliation{Institut de Physique Nucl\'{e}aire IN2P3-CNRS
 and Universit\'{e} Paris-Sud, F-91406 Orsay Cedex, France}
%\affiliation{Universit\'{e} de Lyon 1,
% Institut de Physique Nucl\'{e}aire de Lyon, F-69622 Villeurbanne,
% Cedex, France}


\date{\today}

\begin{abstract}
Applying the semi-realistic $NN$ interactions
that include realistic tensor force
to the Hartree-Fock calculations,
we investigate tensor-force effects on the single-particle levels
in the Ca isotopes.
The semi-realistic interaction successfully describes
the experimental difference
between $\varepsilon(p1s_{1/2})$ and $\varepsilon(p0d_{3/2})$
(denoted by $\mathit{\Delta}\varepsilon_{13}$)
both at $^{40}$Ca and $^{48}$Ca,
confirming importance of the tensor force.
The tensor force plays a role
in the $N$-dependence of $\mathit{\Delta}\varepsilon_{13}$
also in neutron-rich Ca nuclei.
While the $p1s_{1/2}$-$p0d_{3/2}$ inversion is predicted
in heavier Ca nuclei as in $^{48}$Ca,
it takes place only in $N\geq 46$, delayed by the tensor force.
We further investigate possibility of proton bubble structure in Ar,
which is suggested by the $p1s_{1/2}$-$p0d_{3/2}$ inversion
in $^{48}$Ca and more neutron-rich Ca nuclei,
by the spherical Hartree-Fock-Bogolyubov calculations.
Even with the inversion at $^{48}$Ca
the pair correlation prohibits prominent bubble distribution in $^{46}$Ar.
Bubble in Ar is unlikely also near neutron drip line
because either of unboundness or of deformation.
However, $^{34}$Si remains a candidate for proton bubble structure,
owing to large shell gap between $p1s_{1/2}$ and $p0d_{5/2}$.
\end{abstract}

%\keywords{}

\maketitle

%\section{}
%\subsection{}
%\subsubsection{}

% If in two-column mode, this environment will change to single-column
% format so that long equations can be displayed. Use
%\begin{widetext}
%\end{widetext}

%\section{Introduction\label{sec:intro}}
\textit{Introduction.}
Owing to the progress in experiments on unstable nuclei,
it has been recognized~\cite{ref:SP08} that
the nuclear shell structure depends on $Z$ or $N$
as often called \textit{shell evolution}.
Moreover, it is now known~\cite{ref:Vtn,ref:Nak08b,ref:Nak10}
that the tensor force,
which should be contained in the nucleon-nucleon ($NN$) interaction,
plays a crucial role in the shell evolution.
While the tensor force had been ignored
in the conventional mean-field (MF)
or the energy-density-functional (EDF) approaches,
there have been several attempts to incorporate
the tensor force into those approaches;
\textit{e.g.} the calculations
with the Skyrme~\cite{ref:BDO06,ref:CSFB07,ref:BS07,ref:LBB07}
or the Gogny~\cite{ref:OMA06,ref:ACD11} interactions.
However, without well-established strengths and/or radial-dependence
of the tensor force,
it is not straightforward to pin down tensor-force effects quantitatively
from those interactions.
One of the authors (H.N.) has developed
the so-called semi-realistic $NN$ interactions~\cite{ref:Nak03},
which is applicable to the self-consistent MF calculations.
The recent parameter-sets,
M3Y-P$n$ with $n\geq 5$~\cite{ref:Nak08b,ref:Nak10,ref:Nak13},
include the tensor force originating from the $G$-matrix
at the nuclear surface.
Because of the realistic nature of the tensor force in them,
these semi-realistic interactions are suitable
to investigate the tensor-force effects on the shell evolution.

Although the MF or EDF approaches give single-particle (s.p.) levels
in a self-consistent manner,
each of the s.p. levels does not correspond to a single observed state
even beside the doubly magic nuclei,
being fragmented over a certain energy range.
The s.p. energies in the MF approaches should better correspond
to the averaged energy of the states having the specific quantum numbers
weighted by the spectroscopic factors.
There are not many cases in which the spectroscopic factors
have exhaustively been measured with good accuracy.
The proton $1s_{1/2}$ and $0d_{3/2}$ hole states in $^{40}$Ca and $^{48}$Ca
provide us with indispensable examples,
for which sum of the measured spectroscopic factor exceeds
$90\%$~\cite{ref:DWKM76,ref:Ogi87}.
These experimental data have disclosed a notable consequence
that the $p1s_{1/2}$ and $p0d_{3/2}$ levels invert
as $N$ increases from $^{40}$Ca to $^{48}$Ca.
The $p1s_{1/2}$-$p0d_{3/2}$ difference and their inversion
seem to supply a good test of the MF effective interactions (or EDFs).
Since it has been suggested that the tensor force plays a important role
in this inversion~\cite{ref:GMK07,ref:WGZD11},
it is of interest to apply the semi-realistic interaction.
Moreover, the $p1s_{1/2}$-$p0d_{3/2}$ inversion at $^{48}$Ca
suggests a possibility of proton ``bubble'' structure,
depletion of the proton density at the center of the nucleus,
in the two-more-proton deficient nucleus $^{46}$Ar~\cite{ref:KGM08}.
Similar inversion has been predicted
in several Ca nuclei near the neutron drip line,
which suggests proton bubble in Ar.
Possibility of proton bubble structure has been pointed out
also for $^{34}$Si~\cite{ref:GGK09}.
Although it has been difficult to measure charge densities
at the center of unstable nuclei,
the new technology such as SCRIT~\cite{ref:SCRIT}
could open a way to observe such proton bubble structure.
It is noted that correlations beyond MF tend to quench the bubble,
as shown for $^{34}$Si by a recent study
using the generator-coordinate method (GCM)~\cite{ref:Yao12}.
Since the wave function of the GCM ground state was found
spread over a wide range of intrinsic deformations,
it was shown that the level occupation was smeared over the Fermi energy
and the $p1s_{1/2}$ orbital was partially occupied.
These correlations reduce the depletion of the proton density.
However, the transition strengths $\rho^2(E0;0_2^+\rightarrow0_1^+)$
and $B(E2;2_1^+\rightarrow0_1^+)$ are overestimated in Ref.~\cite{ref:Yao12}
compared to the experimental value~\cite{ref:Rot12}.
This discrepancy might indicate
that the two lowest 0$^+$ GCM states are too strongly mixed
in the calculations of Ref.~\cite{ref:Yao12}.

In this paper we apply the self-consistent Hartree-Fock (HF)
and Hartree-Fock-Bogolyubov (HFB) calculations
with the semi-realistic interactions
to the Si to Ca nuclei,
and investigate tensor-force effects
on the shell evolution and the bubble structure.

%\section{Effective Hamiltonian\label{sec:Hamil}}
\textit{Effective Hamiltonian.}
We apply the spherical HF and HFB calculations
by using the Gaussian expansion method~\cite{ref:NS02,ref:Nak06}.
The details of the basis functions are given in Ref.~\cite{ref:Nak08}.
The effective Hamiltonian has the form $H=H_N+V_C-H_\mathrm{c.m.}$,
where $H_N (= \sum_i \mathbf{p}_i^2/2M + \sum_{i<j} v_{ij})$,
$V_C$ and $H_\mathrm{c.m.}$ denote the effective nuclear Hamiltonian,
the Coulomb interaction and the center-of-mass Hamiltonian, respectively.
The exchange term of $V_C$ is treated exactly
and both the one- and the two-body terms of $H_\mathrm{c.m.}$
are subtracted before iteration.

The M3Y-P$n$ semi-realistic interactions have been obtained
by modifying the so-called M3Y-Paris interaction~\cite{ref:M3Y-P},
which was derived by fitting the Yukawa functions
to the $G$-matrix at the nuclear surface.
Density-dependent contact terms have been added to the M3Y-Paris interaction
so as to realize the saturation,
and the LS channels have been enhanced
in order to reproduce the $\ell s$ splitting at the MF level.
We employ the M3Y-P5$'$~\cite{ref:Nak10}
and M3Y-P7~\cite{ref:Nak13} parameter-sets in the following.
In both sets the tensor force of the M3Y-Paris interaction
is maintained without any change,
as in M3Y-P5~\cite{ref:Nak08b} and P6~\cite{ref:Nak13}.
For comparison we also use the D1M~\cite{ref:D1M} parameter-set
of the Gogny interaction,
which has no tensor channels.
Whereas we have implemented calculations with M3Y-P6 and D1S~\cite{ref:D1S},
results of M3Y-P6 (D1S) are similar to those of M3Y-P7 (D1M).

\begin{figure}
\includegraphics[scale=1.0]{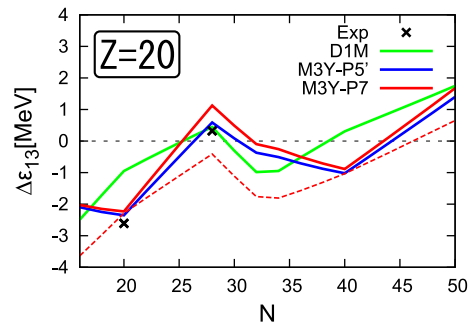}
\caption{(Color) $\mathit{\Delta}\varepsilon_{13}$ of the Ca isotopes.
Green, blue and red lines represent the results
with the D1M, M3Y-P5$'$ and P7 interactions, respectively.
Thin red dashed line is obtained from M3Y-P7
but by removing contribution of the tensor force.
\label{fig:de13}}
\end{figure}

%\section{$N$-dependence of $p1s_{1/2}$-$p0d_{3/2}$ levels in $Z=20$ nuclei
%\label{sec:Ca-sp}}
\textit{$N$-dependence of $p1s_{1/2}$-$p0d_{3/2}$ levels in $Z=20$ nuclei.}
We here express the s.p. energy difference under interest as
$\mathit{\Delta}\varepsilon_{13}
= \varepsilon(p1s_{1/2})-\varepsilon(p0d_{3/2})$.
Figure~\ref{fig:de13} shows $N$-dependence of $\mathit{\Delta}\varepsilon_{13}$
in the Ca isotopes
obtained by the spherical HF calculations.
The experimental values of $\mathit{\Delta}\varepsilon_{13}$
in $^{40}$Ca and $^{48}$Ca are also displayed for comparison,
which are obtained after average
weighted by the spectroscopic factors~\cite{ref:DWKM76,ref:Ogi87}.
Although significant $N$-dependence is found in the experiments
on $\mathit{\Delta}\varepsilon_{13}$ of the Ca isotopes,
not many MF interactions (or EDFs) can reproduce this $N$-dependence
quantitatively~\cite{ref:GMK07,ref:WGZD11}.
In practice, despite agreement of $\mathit{\Delta}\varepsilon_{13}$
with the data at $^{48}$Ca,
there is significant discrepancy at $^{40}$Ca in the D1M result,
as viewed in Fig.~\ref{fig:de13}.
With D1S, $\mathit{\Delta}\varepsilon_{13}$ is slightly shifted downward
and the inversion at $^{48}$Ca does not occur.
The slope of $\mathit{\Delta}\varepsilon_{13}$ from $^{40}$Ca to $^{48}$Ca
in the D1M result is typical to the MF calculations with no tensor force.
On the contrary, as depicted in Fig.~\ref{fig:de13},
the semi-realistic M3Y-P5$'$ and P7 interactions successfully reproduce
the $N$-dependence of $\mathit{\Delta}\varepsilon_{13}$.
In Fig.~\ref{fig:de13} we also show $\mathit{\Delta}\varepsilon_{13}$
in which contribution of the tensor force is removed from the M3Y-P7 result.
Then $\mathit{\Delta}\varepsilon_{13}$ varies in parallel to those of D1M,
unable to reproduce the observed slope.
This confirms the crucial role of the tensor force
in $\mathit{\Delta}\varepsilon_{13}$.

It has been suggested~\cite{ref:SP13} that,
although the experimental information of $\varepsilon(p0d_{5/2})$
is not complete,
the $\ell s$ splitting $\mathit{\Delta}\varepsilon_{35}
= \varepsilon(p0d_{3/2})-\varepsilon(p0d_{5/2})$ depends on $N$ as well,
decreasing from $^{40}$Ca to $^{48}$Ca by $1.8\,\mathrm{MeV}$.
Hardly described by the Gogny interaction
without the tensor force~\cite{ref:SP13},
this reduction of the $\ell s$ splitting may be accounted for
as a role of the tensor force~\cite{ref:UOB12}.
We have reduction of $\mathit{\Delta}\varepsilon_{35}$
from $^{40}$Ca to $^{48}$Ca
by $3.5\,\mathrm{MeV}$ ($3.1\,\mathrm{MeV}$) with M3Y-P7 (P5$'$).

The variation of $\mathit{\Delta}\varepsilon_{13}$ from $N=20$ to $28$
is a result of occupation of the $n0f_{7/2}$ orbit.
As $n0f_{7/2}$ is occupied,
the tensor force tends to lower $p0d_{3/2}$ but not $p1s_{1/2}$,
therefore increasing $\mathit{\Delta}\varepsilon_{13}$.
In D1M the parameters in the central and LS channels
have globally been adjusted to experimental data,
and the inversion of the s.p. levels at $^{48}$Ca is reproduced.
However, because lacking the tensor force, this inversion takes place
at the expense of the discrepancy in $\mathit{\Delta}\varepsilon_{13}$
at $^{40}$Ca.
As mentioned above, most effective interactions investigated so far,
even including phenomenological tensor channels, have failed to reproduce
the slope of $\mathit{\Delta}\varepsilon_{13}$~\cite{ref:GMK07,ref:WGZD11},
with the only exception of SLy5+$T_w$.
It is remarked that the semi-realistic interactions reproduce the slope,
owing to the realistic tensor force,
as well as the absolute values.
The SLy5+$T_w$ interaction~\cite{ref:Bai10},
in which the strength parameters of the zero-range tensor force
are determined from the $G$-matrix~\cite{ref:SBF77},
gives appropriate slope of $\mathit{\Delta}\varepsilon_{13}$~\cite{ref:WGZD11}
in $^{40-48}$Ca.
However, the absolute values of $\mathit{\Delta}\varepsilon_{13}$
in the SLy5+$T_w$ results are substantially higher than the data.
It should be commented here that
there might be influence of ground-state correlations
on the s.p. levels beyond fragmentation,
and that its assessment is desired for complete understanding.

The tensor force affects $\mathit{\Delta}\varepsilon_{13}$
from $N=34$ to $40$, \textit{i.e.} as $n0f_{5/2}$ is occupied.
Experiments on the $p1s_{1/2}$ and the $p0d_{3/2}$ hole states around $^{60}$Ca
may provide further evidence of the tensor-force effects,
if carried out in the future.
Although it depends on the effective interactions
where the neutron drip line of Ca is,
$p1s_{1/2}$-$p0d_{3/2}$ inversion could occur again toward $^{70}$Ca
as $n0g_{9/2}$ is occupied.
However, whereas the inversion takes place already at $^{60}$Ca
with D1M,
the inversion is delayed until $N=46$ in the M3Y-P5$'$ and P7 results,
in which the parameters have been determined
in the presence of the tensor force.
Comparing the present results for $\mathit{\Delta}\varepsilon_{13}$
to those of the phenomenological tensor channels~\cite{ref:WGZD11},
we do not view quantitative agreement in $^{48-70}$Ca,
including SLy5+$T_w$.
Since it is based on the realistic tensor force
that well reproduces $\mathit{\Delta}\varepsilon_{13}$ in $^{40-48}$Ca
without adjustment,
the prediction by the semi-realistic interactions
seems reliable also in $^{48-70}$Ca.

%\section{Investigation of proton bubbles\label{sec:bubble}}
\textit{Investigation of proton bubbles.}
The $p1s_{1/2}$-$p0d_{3/2}$ inversion in $^{48}$Ca suggests
that dominant configuration of the two-proton-deficient nucleus $^{46}$Ar
is $(p1s_{1/2})^{-2}$.
Since only the $s$-states give sizable density at the center of nuclei,
$^{46}$Ar is a candidate of a nucleus
having proton bubble structure~\cite{ref:KGM08}.
While it is not easy to measure matter or neutron densities
in a model-independent manner particularly for unstable nuclei,
charge densities may unambiguously be extracted
from the electron scattering experiments.
Since the M3Y-P$n$ semi-realistic interactions successfully
describe the $N$-dependence of $\mathit{\Delta}\varepsilon_{13}$,
indicating that mechanism giving rise to the $p1s_{1/2}$-$p0d_{3/2}$ inversion
is correctly contained,
it will be of certain interest to investigate
possibility of proton bubble structure by applying these interactions.

\begin{figure}
\includegraphics[scale=1.0]{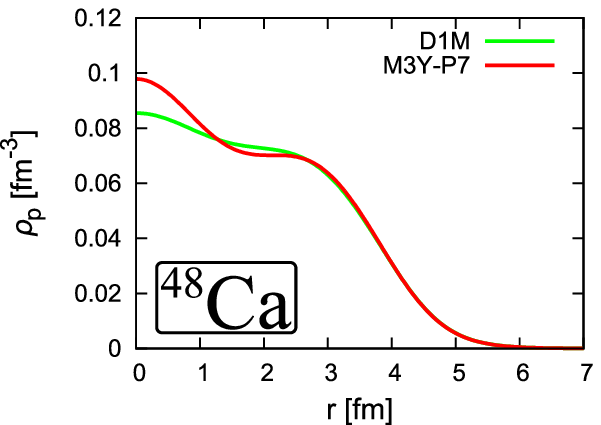}\\
\includegraphics[scale=1.0]{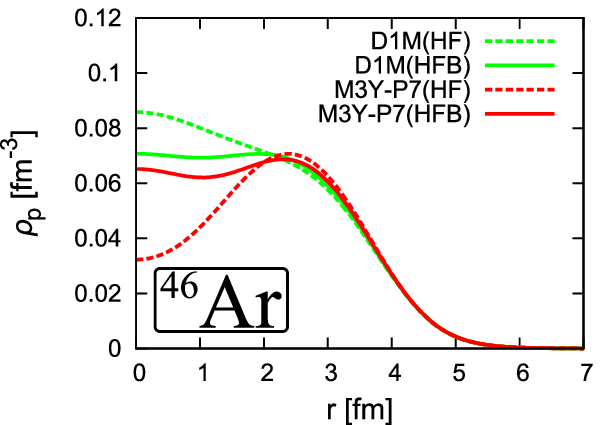}\\
\caption{(Color) Proton density distributions in $^{48}$Ca and $^{46}$Ar
 obtained from the HF and HFB calculations.
\label{fig:rho_Ar46}}
\end{figure}

We present proton density distributions at $^{48}$Ca and $^{46}$Ar obtained
from the MF calculations with D1M and M3Y-P7 in Fig.~\ref{fig:rho_Ar46}.
The pair correlation is quenched in the ground-state of $^{48}$Ca
and the HF and HFB results are identical.
Within the spherical HF regime with M3Y-P7,
we have depletion of the proton distribution around the origin at $^{46}$Ar,
since $p1s_{1/2}$ becomes unoccupied if compared with $^{48}$Ca.
The same holds for M3Y-P5$'$, though not displayed.
Such depletion is not found in the HF result with D1M,
in which the ground state has the $(p0d_{3/2})^{-2}$ configuration.
Despite the $p1s_{1/2}$-$p0d_{3/2}$ inversion at $^{48}$Ca,
the total energy of the $(p0d_{3/2})^{-2}$ state becomes lower
than that of the $(p1s_{1/2})^{-2}$ state at $^{46}$Ar in the D1M result.
On the other hand, once the pair correlation is taken into account,
small energy difference between $p1s_{1/2}$ and $p0d_{3/2}$
significantly mixes up the $(p1s_{1/2})^{-2}$
and the $(p0d_{3/2})^{-2}$ configurations for both interactions.
Thus the depletion observed in the HF densities with M3Y-P7
is smoothed out when the pairing is switched on.
Notice that this consequence differs from that derived
by SLy5+$T_w$~\cite{ref:WGZD11}.
This may be because SLy5+$T_w$ gives larger $\mathit{\Delta}\varepsilon_{13}$
in $^{48}$Ca than the M3Y-P$n$ interactions and the data,
though reproducing the slope in $^{40-48}$Ca.

While one might expect the formation of proton bubbles around $^{60}$Ca
because of the $p1s_{1/2}$-$p0d_{3/2}$ inversion given by D1M,
the tensor force prohibits the inversion
and accordingly the proton bubble structure,
as indicated by Fig.~\ref{fig:de13}.
As previously discussed this is mostly due to the occupation of $n0f_{5/2}$.
In contrast, the inversion is predicted
by all the interactions under investigation around $^{70}$Ca.
%With the D1S interaction, however, $^{70}$Ca is out of the neutron drip line.
The $^{70}$Ca nucleus is bound
in the HFB calculations with D1M, M3Y-P5$'$
and P7~\cite{ref:Nak13,ref:Nak10b}.
At a glance,
the $p1s_{1/2}$-$p0d_{3/2}$ inversion exhibited in Fig.~\ref{fig:de13}
seems to suggest proton bubble structure in $^{64-68}$Ar even with M3Y-P7.
However, the $^{64-68}$Ar nuclei lie beyond the neutron-drip line
while $^{66-70}$Ca are bound,
within the spherical HFB calculation.
Although these Ar nuclei might be deformed
and thereby their energy could be lowered enough to be bound,
the deformation significantly mixes the $(p1s_{1/2})^{-2}$ state with others,
which easily destroys bubble structure.
We therefore conclude that proton bubble structure is unlikely
to be observed in any of the Ar nuclei.

\begin{figure}
\includegraphics[scale=1.0]{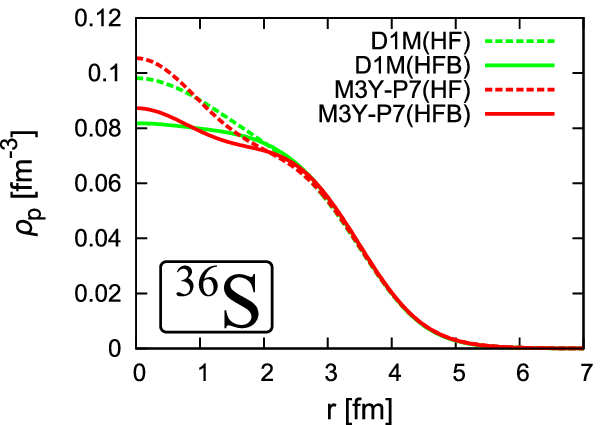}\\
\includegraphics[scale=1.0]{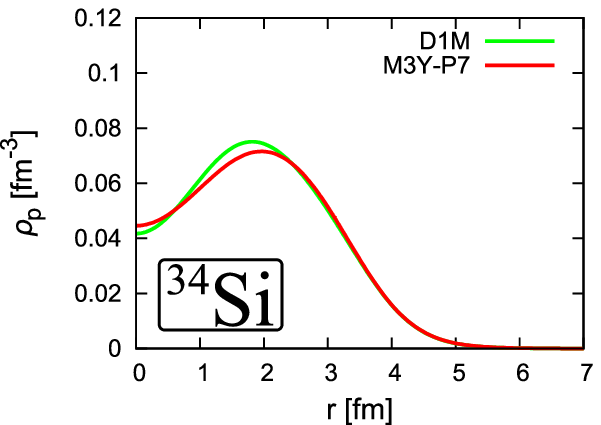}
\caption{(Color) Proton density distributions in $^{36}$S and $^{34}$Si
 obtained from the HF and HFB calculations.
\label{fig:rho_Si34}}
\end{figure}

While the $N=20$ magic number disappears in the proton-deficient region
of $Z\leq 12$,
it remains magic in $Z\geq 14$, keeping the nuclear shape spherical.
Possibility of proton bubble structure has been pointed out
also for $^{34}$Si~\cite{ref:GGK09}.
The energy difference $\mathit{\Delta}\varepsilon_{15}
= \varepsilon(p1s_{1/2})-\varepsilon(p0d_{5/2})$
exceeds $5\,\mathrm{MeV}$ for $^{36}$S in the spherical HF calculations,
irrespective of the effective interactions under consideration.
We note that $-\mathit{\Delta}\varepsilon_{13}$
is less than $2\,\mathrm{MeV}$,
giving rise to sizable pair excitation at $^{36}$S.
The occupation probability of $p1s_{1/2}$ is $0.6-0.7$ in the HFB results.
However, owing to the large subshell gap $\mathit{\Delta}\varepsilon_{15}$,
the ground state of $^{34}$Si is expected to be $(p1s_{1/2})^{-2}$
with good approximation.
Proton density distribution of $^{34}$Si is depicted
in Fig.~\ref{fig:rho_Si34} in comparison with that of $^{36}$S.
Since $\mathit{\Delta}\varepsilon_{15}$ is sufficiently large,
prominent proton bubble structure is predicted in $^{34}$Si
in the MF calculations for all the effective interactions
considered in this paper.
The large $\mathit{\Delta}\varepsilon_{15}$
prevents the pair correlation from mixing the $(p1s_{1/2})^{-2}$ state
with the other states (\textit{e.g.} $(p0d_{5/2})^{-2}$)
in the ground state of $^{34}$Si,
giving identical density distribution between HF and HFB.
It has been shown~\cite{ref:Yao12} that correlations beyond the MF regime
substantially weaken the central depletion of the proton density in $^{34}$Si.
The bubble may become less conspicuous by the finite-size effect of protons,
as well.
However, degree of the smearing due to the correlations is not obvious,
and there remains possibility for the proton bubble to survive in $^{34}$Si.
Future experiments on the charge density of this nucleus are awaited.

%\section{Summary\label{sec:summary}}
\textit{Summary.}
We have investigated tensor-force effects
on the shell evolution and the bubble structure
around the $Z$ or $N=20$ magic number,
by applying the self-consistent HF and HFB calculations
with the semi-realistic interactions.
We have shown that the realistic tensor force gives adequate $N$-dependence
of the $p1s_{1/2}$-$p0d_{3/2}$ difference in the Ca isotopes,
if combined with appropriate central and LS forces
as in the M3Y-P$n$ semi-realistic interactions.
Although the $p1s_{1/2}$-$p0d_{3/2}$ inversion is predicted
for the Ca nuclei near the neutron drip line,
the inversion is delayed with the interactions
including realistic tensor force,
suggesting the inversion only in $^{66-70}$Ca.

On the basis that semi-realistic interactions
reproduce the observed $p1s_{1/2}$-$p0d_{3/2}$ inversion reasonably well,
we apply them to the proton-bubble-structure problem.
In $^{46}$Ar, although we view depletion of the proton density within HF,
it is predicted that pair correlation prohibits the bubble structure
from being realized.
Although one may anticipate proton bubble structure
in the highly neutron-excess nuclei $^{64-68}$Ar
because of the inversion in $^{66-70}$Ca,
these nuclei might be unbound if not deformed,
and therefore proton bubble structure is unlikely.
On the contrary, the possibility of the proton bubble structure
is not ruled out for $^{34}$Si,
though there should be correlation effects that will make
the central depletion of the proton density less conspicuous
than the MF prediction.

~
The authors are grateful to P.-H. Heenen and O. Sorlin
for valuable discussion and comments.
This work is financially supported as Grant-in-Aid for Scientific Research
Nos.~22540266 and 25400245 by JSPS, and No.~24105008 by MEXT.
Numerical calculations are performed on HITAC SR16000s
at IMIT in Chiba University, at YITP in Kyoto University,
at ITC in University of Tokyo, and at IIC in Hokkaido University.
%\end{acknowledgments}


\end{document}